%% file: manuscript.tex
\documentclass[final]{svjour3}

\usepackage{amssymb}
\usepackage{graphicx}
\usepackage{mathptmx}
\usepackage{multirow}
\usepackage[numbers]{natbib}

\makeatletter
\journalname{Journal of Low Temperature Physics}

\makeatletter
\def\input@path{{./}{sections/}}
\makeatother
\graphicspath{{./}{figures/}}

\begin{document}
    \input{frontmatter}
    \newpage
    \input{abstract}
    \input{introduction}
    \input{deshima-2.0}
    \input{instrument}
    \input{observation}
    \input{science}
    \input{conclusions}
    \input{backmatter}
\end{document}

%% file: frontmatter.tex
% !TEX root = ../manuscript.tex
\title{DESHIMA 2.0: development of an integrated superconducting spectrometer for science-grade astronomical observations}

\author{
    Akio Taniguchi$^{1,a}$
    \and Tom J. L. C. Bakx$^{1,2}$
    \and Jochem J. A. Baselmans$^{3,4}$
    \and Robert Huiting$^{4}$
    \and Kenichi Karatsu$^{3,4}$
    \and Nuria Llombart$^{3}$
    \and Matus Rybak$^{3,5}$
    \and Tatsuya Takekoshi$^{6,7}$
    \and Yoichi Tamura$^{1}$
    \and Hiroki Akamatsu$^{4}$
    \and Stefanie Brackenhoff$^{3,8}$
    \and Juan Bueno$^{3}$
    \and Bruno T. Buijtendorp$^{3}$
    \and Shahab Dabironezare$^{3}$
    \and Anne-Kee Doing$^{3}$
    \and Yasunori Fujii$^{2}$
    \and Kazuyuki Fujita$^{9}$
    \and Matthijs Gouwerok$^{3}$
    \and Sebastian H\"{a}hnle$^{3,4}$
    \and Tsuyoshi Ishida$^{7}$
    \and Shun Ishii$^{2,10}$
    \and Ryohei Kawabe$^{2,11}$
    \and Tetsu Kitayama$^{12}$
    \and Kotaro Kohno$^{7,13}$
    \and Akira Kouchi$^{9}$
    \and Jun Maekawa$^{2}$
    \and Keiichi Matsuda$^{1}$
    \and Vignesh Murugesan$^{4}$
    \and Shunichi Nakatsubo$^{14}$
    \and Tai Oshima$^{2,11}$
    \and Alejandro Pascual Laguna$^{3,4}$
    \and David J. Thoen$^{3}$
    \and Paul P. van der Werf$^{5}$
    \and Stephen J. C. Yates$^{15}$
    \and Akira Endo$^{3}$
}

\institute{
    a. \email{taniguchi@a.phys.nagoya-u.ac.jp}\\
    1. Division of Particle and Astrophysical Science, Graduate School of Science, Nagoya University, Furocho, Chikusa-ku, Nagoya, Aichi 464-8602, Japan\\
    2. National Astronomical Observatory of Japan, 2-21-1 Osawa, Mitaka, Tokyo 181-8588, Japan\\
    3. Faculty of Electrical Engineering, Mathematics and Computer Science, Delft University of Technology, Mekelweg 4, 2628 CD Delft, The Netherlands\\
    4. SRON---Netherlands Institute for Space Research, Niels Bohrweg 4, 2333 CA Leiden, The Netherlands\\
    5. Leiden Observatory, Leiden University, PO Box 9513, 2300 RA Leiden, The Netherlands\\
    6. Kitami Institute of Technology, 165 Koen-cho, Kitami, Hokkaido 090-8507, Japan\\
    7. Institute of Astronomy, Graduate School of Science, The University of Tokyo, 2-21-1 Osawa, Mitaka, Tokyo 181-0015, Japan\\
    8. Kapteyn Astronomical Institute, University of Groningen, PO Box 800, NL-9700 AV Groningen, The Netherlands\\
    9. Institute of Low Temperature Science, Hokkaido University, Sapporo 060‑0819, Japan\\
    10. Joint ALMA Observatory, Alonso de C\'{o}rdova 3107, Vitacura, Santiago, Chile\\
    11. The Graduate University for Advanced Studies (SOKENDAI), 2-21-1 Osawa, Mitaka, Tokyo 181-0015, Japan\\
    12. Department of Physics, Toho University, 2-2-1 Miyama, Funabashi, Chiba 274-8510, Japan\\
    13. Research Center for the Early Universe, Graduate School of Science, The University of Tokyo, 7‑3‑1 Hongo, Bunkyo‑ku, Tokyo 113‑0033, Japan\\
    14. Institute of Space and Astronautical Science, Japan Aerospace Exploration Agency, Sagamihara 252‑5210, Japan\\
    15. SRON---Netherlands Institute for Space Research, Landleven 12, 9747 AD Groningen, The Netherlands
}

\titlerunning{Development of DESHIMA 2.0}
\authorrunning{Taniguchi et al.}

\maketitle

%% file: abstract.tex
% !TEX root = ../manuscript.tex
\begin{abstract}

Integrated superconducting spectrometer (ISS) technology will enable ultra-wideband, integral-field spectroscopy for (sub)millimeter-wave astronomy, in particular, for uncovering the dust-obscured cosmic star formation and galaxy evolution over cosmic time.
Here we present the development of DESHIMA 2.0, an ISS for ultra-wideband spectroscopy toward high-redshift galaxies.
DESHIMA 2.0 is designed to observe the 220--440 GHz band in a single shot, corresponding to a redshift range of $z=3.3\textrm{--}7.6$ for the ionized carbon emission ([C II] 158~$\mu$m).
The first-light experiment of DESHIMA 1.0, using the 332--377 GHz band, has shown an excellent agreement among the on-sky measurements, the lab measurements, and the design.
As a successor to DESHIMA 1.0, we plan the commissioning and the scientific observation campaign of DESHIMA 2.0 on the ASTE 10-m telescope in 2023.
Ongoing upgrades for the full octave-bandwidth system include the wideband 347-channel chip design and the wideband quasi-optical system.
For efficient measurements, we also develop the observation strategy using the mechanical fast sky-position chopper and the sky-noise removal technique based on a novel data-scientific approach.
In the paper, we show the recent status of the upgrades and the plans for the scientific observation campaign.


\end{abstract}

%% file: introduction.tex
% !TEX root = ../manuscript.tex
\section{Introduction}
\label{s:introduction}

Wideband spectroscopy at millimeter and submillimeter wavelengths is a promising approach to uncovering the star formation history and galaxy evolution in the early universe.
With the next-generation large single-dish telescopes \cite{Kawabe_2016, Klaassen_2020, Lou_2020}, it offers a unique opportunity to probe the physical and chemical properties of dust-obscured interstellar media using atomic and molecular emission lines.
For this purpose, the concept of an integrated superconducting spectrometer (ISS) has been proposed, which enables ultra-wideband ($\gtrsim 100$~GHz) and medium-resolution ($F/\Delta F \sim 500$) spectroscopy based on an on-chip filtering circuit \cite{Wheeler_2016, Endo_2019a} or an on-chip diffraction grating \cite{Cataldo_2018}, and microwave kinetic inductance detectors (MKIDs).

The deep spectroscopic high-redshift mapper (DESHIMA \cite{Endo_2019a}) is an ISS primarily designed for blind redshift identification of dusty star-forming galaxies (DSFGs) in the $z\gtrsim 3$ universe using bright emission lines (e.g., [C~II] 158~$\mu$m, [O~III] 88~$\mu$m).
The goal of DESHIMA is to achieve an instantaneous spectral bandwidth of $\sim 200$~GHz in the 1-mm band, which corresponds to the detectable redshift range of $z\sim 3\textrm{--}7$ in the case of [C~II] observations.

We gave the first on-sky demonstration of ISS using a prototype DESHIMA that covered a 45-GHz (332--377~GHz) bandwidth (referred to as DESHIMA 1.0) on the Atacama Submillimeter Telescope Experiment (ASTE) 10-m telescope \cite{Ezawa_2004, Ezawa_2008} in 2017.
We detected molecular emission lines from both Galactic and extragalactic sources.
The noise equivalent flux density (NEFD) reached in these successful observations was consistent with that predicted from a theoretical model\footnote{A model that includes only photon noise and quasi-particle recombination noise as noise sources.}, showing an excellent agreement among the on-sky measurements, the lab measurements and the design \cite{Endo_2019b}.

In this paper, we present the development of DESHIMA 2.0, a successor to DESHIMA 1.0, for science-grade high-redshift spectroscopy.
We overview DESHIMA 2.0 (Section~\ref{s:deshima-2.0}) and show the upgrades of the instrument with an initial laboratory measurement (Section~\ref{s:upgrades-of-the-instrument}).
We describe the observation strategy for the detection of faint sources (Section~\ref{s:upgrades-of-observation-strategy}).
Finally we introduce the possible science cases of DESHIMA 2.0 on ASTE (Section~\ref{s:science-cases}).

%% file: deshima-2.0.tex
% !TEX root = ../manuscript.tex
\section{DESHIMA 2.0}
\label{s:deshima-2.0}

\begin{figure*}[t]
    \centering
    \includegraphics[width=\textwidth]{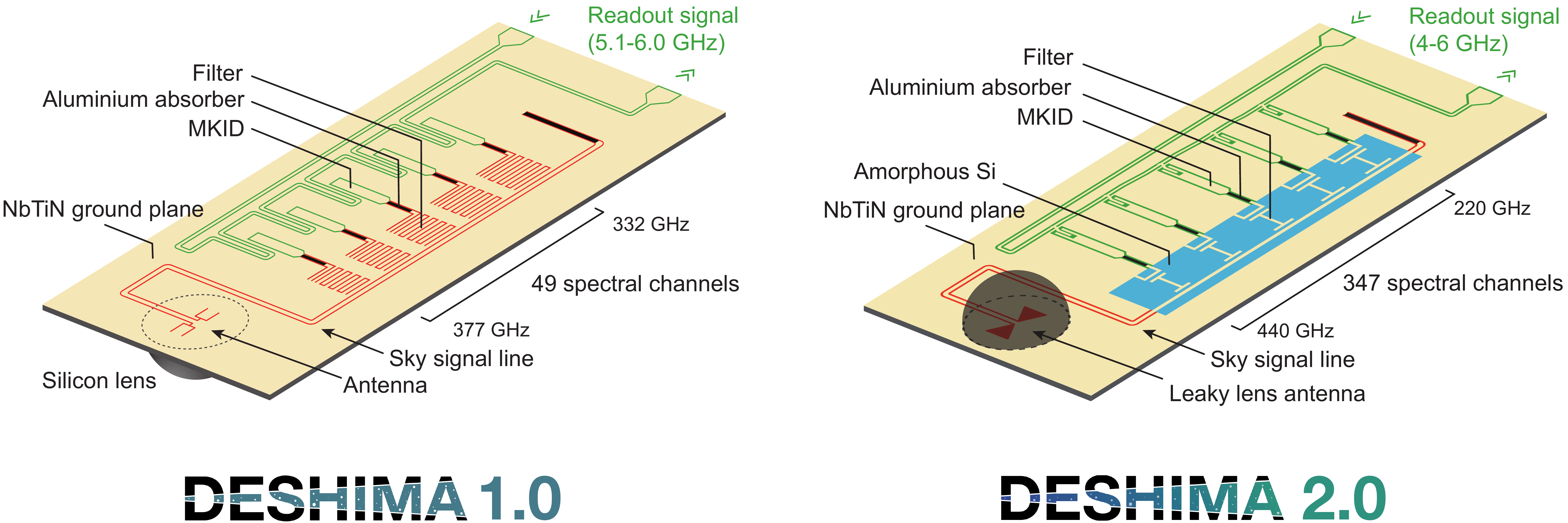}
    \caption{
        The chip design of DESHIMA 2.0 (right panel) in comparison with that of DESHIMA 1.0 \cite{Endo_2019a, Endo_2019b} (left panel).
        The sky signal including both astronomical and atmospheric emission is coupled by an ultra-wideband leaky-lens antenna \cite{Dabironezare_2020} and is guided to the subsequent filterbank.
        Each filter connects with an MKID and the change in the resonance frequency by incoming photons is measured through a single readout signal.
        An offline calibration is then carried out to convert the measured filter response into the brightness temperature of the signal \cite{Takekoshi_2020}.
        The atmospheric emission is removed by offline signal processing (see also Section~\ref{subs:the-data-scientific-noise-removal-technique}).
    }
    \label{fig:chip-design}
\end{figure*}

DESHIMA 2.0 is a single-pixel ISS operating at submillimeter wavelength and offers an instantaneous spectral bandwidth of 220~GHz ranging 220--440~GHz with 347 spectral channels.
Fig.~\ref{fig:chip-design} shows the chip design of DESHIMA 2.0 in comparison with that of DESHIMA 1.0.
DESHIMA 2.0 is designed to be installed on the ASTE 10-m telescope.
The installation and commissioning of DESHIMA 2.0 and a subsequent three-month scientific observation campaign are planned in 2023.

One of the goals of DESHIMA 2.0 is to detect the redshifted [C~II] emission from a bright DSFG (observed infrared luminosity of $L_{\mathrm{IR}}\gtrsim10^{13}L_{\odot}$) with an eight-hour (i.e., a night) ASTE observation including any overheads.
The expected line flux ($\sim 10^{-18}$~W~m$^{-2}$) is, however, fainter by at least an order of magnitude than that of the faintest target of DESHIMA 1.0.
To achieve this goal, we upgrade both the instrument design and the observation strategy, the details of which are described in Sections~\ref{s:upgrades-of-the-instrument}~and~\ref{s:upgrades-of-observation-strategy}, respectively.

\begin{table*}[h]
    \centering
    \begin{tabular}{llll}
        \hline\hline
        ~ & Specification & DESHIMA 1.0 & DESHIMA 2.0\\
        \hline
        \multirow{8}{*}{Instrument} & Observed frequency range (GHz) & 332--377 & 220--440\\
        ~ & (Detectable redshift range of [C~II]) & 4.0--4.7 & 3.3--7.6\\
        ~ & Number of spectral channels & 49 & 347\\
        ~ & Number of spatial pixels & 1 & 1\\
        ~ & Number of polarizations & 1 & 1\\
        ~ & Mean frequency resolution ($F/\Delta F$) & 380 & 500\\
        ~ & Instrument optical efficiency (\%) & 2 & 8 (baseline), 16 (goal)\\
        ~ & Antenna design & double-slot & leaky-lens\\
        \hline
        \multirow{2}{*}{Observation} & On-source fraction (\%) & 8 & 30 (baseline), 40 (goal)\\
        ~ & Noise-removal method & direct subtraction & data-scientific subtraction\\
        \hline
    \end{tabular}
    \caption{
        Summary of the capabilities of DESHIMA 2.0 in comparison with DESHIMA 1.0.
        Note that we show both baseline (minimum requirement) and goal values in each specification of DESHIMA 2.0.
    }
    \label{tab:summary-of-upgrades}
\end{table*}

\begin{figure*}[t]
    \centering
    \includegraphics[width=\textwidth]{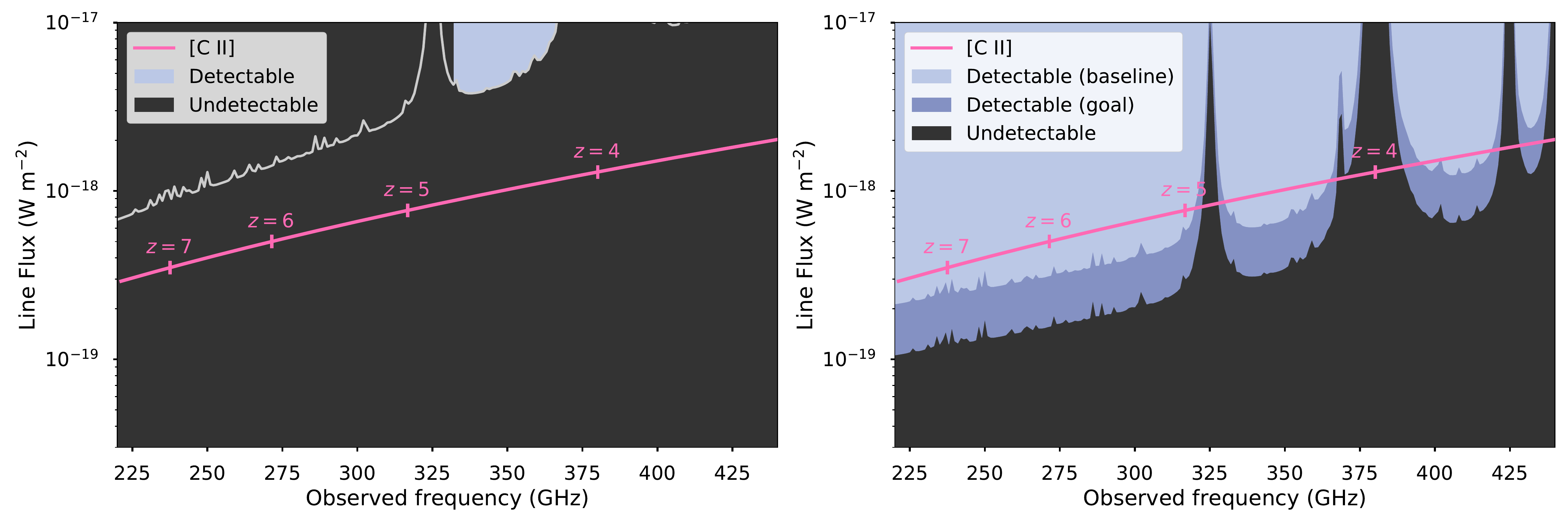}
    \caption{
        Observed-frequency and line-flux space (colored regions) within which an astronomical signal can be detected at S/N$>$5 with an eight-hour ASTE observation of DESHIMA 1.0 (left panel) and DESHIMA 2.0 (right panel).
        Overhead time such as off-source measurements is included in the observation time.
        An antenna elevation of 60~degree and a zenith precipitable water vapor of 1.0~mm are assumed during the observation.
        In both panels, we show the expected [C~II] line flux curves of a DSFG ($L_{\mathrm{IR}} = 3 \times 10^{13} L_{\odot}$) as a function of redshift ($z=3.3\textrm{--}7.6$), assuming the infrared-to-[C~II] conversion factor of $10^{-2.74}$ \cite{Bonato_2014}.
        Note that there exist strong atmospheric absorptions at around 325, 365, 380, 425, and 440~GHz, where the detection needs much longer observation times or is impossible.
        The values are estimated by the open-source code \texttt{deshima-sensitivity} \cite{deshima-sensitivity}.
    }
    \label{fig:detectable-line-flux}
\end{figure*}

Table~\ref{tab:summary-of-upgrades} and Fig.~\ref{fig:detectable-line-flux} show the capabilities and the expected sensitivity of DESHIMA 2.0 in comparison with DESHIMA 1.0, respectively.
The observed frequency range corresponds to the detectable redshift range of $z=3.3\textrm{--}7.6$ for [C~II].
With improved sensitivity and efficiency, [C~II] observations with DESHIMA 2.0 will bring us spectroscopic (i.e., redshift) identification of a single bright DSFG per night, which is efficient even compared with those with Atacama Large Millimeter/submillimeter Array (ALMA \cite{Wootten_2009}).

%% file: instrument.tex
% !TEX root = ../manuscript.tex
\section{Upgrades of the instrument}
\label{s:upgrades-of-the-instrument}

\subsection{The chip design}
\label{subs:the-chip-design}

To contiguously cover the 220~GHz bandwidth, we increase the number of filters and optimize the frequency resolution ($F/\Delta F$) of the DESHIMA 2.0 chip.
We design to have 347 spectral channels and $F/\Delta F \simeq 500$ across the 220~GHz bandwidth \cite{PascualLaguna_2021}.
Combined with the significant improvement in the instrument optical efficiency (see also table~\ref{tab:summary-of-upgrades}), the DESHIMA 2.0 chip aims to cover more than 90\% of the bandwidth with sufficient filter response.

Fig.~\ref{fig:deshima-2.0-chip} shows the first-fabricated DESHIMA 2.0 chip and a laboratory measurement of the MKID response, demonstrating the instantaneous coverage of 222--425~GHz.
Although the center frequencies and the Q factors of some MKIDs still require to be optimized in future fabrication runs, the current chip already merits telescope observations.
The feasible sensitivity taking these effects into account are shown in a separate paper \cite{Rybak_2022}.
The fabrication of the on-chip filterbank will be discussed in detail in \cite{Thoen_2022}.

\subsection{The quasi-optics design}
\label{subs:the-quasi-optics-design}

\begin{figure*}[t]
    \centering
    \includegraphics[width=\textwidth]{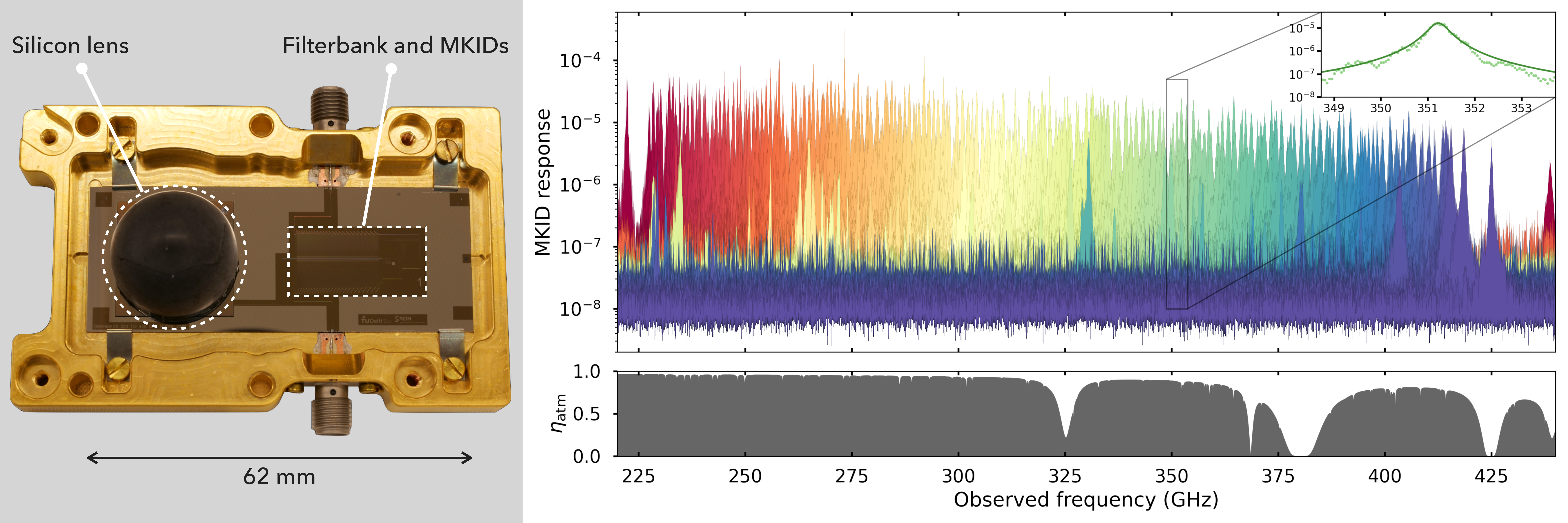}
    \caption{
        A photograph of the first-fabricated DESHIMA 2.0 chip in a holder (left panel) and a laboratory measurement of the MKID response of each filter (right panel; in relative frequency shift, $\delta f / f$ \cite{Takekoshi_2020}), plotted with the atmospheric transmission under the same conditions as Fig.~\ref{fig:detectable-line-flux}.
        The number of spectral filters is 285, and the mean and standard deviation of their Q factors are 690 and 260, respectively.
        The inset shows the measurement of a typical spectral filter (center frequency of $\sim$350 GHz) with a Lorentzian curve fit overlaid.
        Note that the slight filter gaps seen at 225 and 420~GHz are intentional for the laboratory measurement (i.e., making most outer filters isolated from others).
    }
    \label{fig:deshima-2.0-chip}
\end{figure*}

To achieve good coupling efficiencies across an octave bandwidth, we adopt the leaky-lens antenna design \cite{Neto_2010a, Neto_2010b} for DESHIMA 2.0.
Unlike the previous resonant double-slot antenna, it offers frequency-independent beams and high aperture efficiency at submillimeter wavelength \cite{Hahnle_2020}.
Fig.~\ref{fig:deshima-2.0-chip} also shows a silicon lens that covers the leaky-wave slot.

We estimate the aperture efficiency of the ASTE-DESHIMA system based on the leaky-len antenna and the cryogenic system for DESHIMA 2.0.
A simulation demonstrates that $\eta_{\mathrm{ap}} > 0.55$ can be achieved over the entire frequency range of DESHIMA 2.0 \cite{Dabironezare_2020}.

%% file: observation.tex
% !TEX root = ../manuscript.tex
\section{Upgrades of observation strategy}
\label{s:upgrades-of-observation-strategy}

\subsection{The fast sky-position chopper}
\label{subs:the-fast-sky-position-chopper}

As a ground-based telescope instrument, removing fluctuation noises of the sky emission and the instrument (e.g., two-level noise) from observed data is essential for DESHIMA.
We employ the widely-used position-switching (PSW) method, where a sky reference (off-source) spectrum is subtracted from a target (on-source) spectrum.
The on-source efficiency is, however, often much lower than 50\% due to the large dead time to move between two sky positions by antenna drive (see also table~\ref{tab:summary-of-upgrades} for the case of DESHIMA 1.0).

To improve the efficiency, we develop the sky-position chopper designed to be installed between the secondary and tertiary mirrors of ASTE.
Fig.~\ref{fig:observation-strategy} shows its illustration.
Without antenna drive, it enables to switch two optical paths that point to different sky positions 234 arcsec apart.
The switching frequency is set at 10~Hz so as to be faster than the knee frequencies of both the atmospheric noise and the dielectric two-level-system (TLS) noise ($\sim 1$~Hz \cite{Huijten_2022}).
We estimate that the efficiency reaches 30--40\%, which approaches the theoretical maximum on-source efficiency of 50\%.

\subsection{The data-scientific noise removal technique}
\label{subs:the-data-scientific-noise-removal-technique}

As expected flux densities of scientific targets, especially DSFGs, are 3--4 orders of magnitudes fainter than the sky noise, longer ($\gtrsim 10$~hr) observations are often required to detect the line and continuum signals.
Here, another issue of the PSW method is the degradation of the achieved noise level by a factor of $\sqrt{2}$ caused by direct subtraction between noisy spectra; in other words, it would take twice as long time as the pure photon-noise limit.

\begin{figure*}[t]
    \centering
    \includegraphics[width=\textwidth]{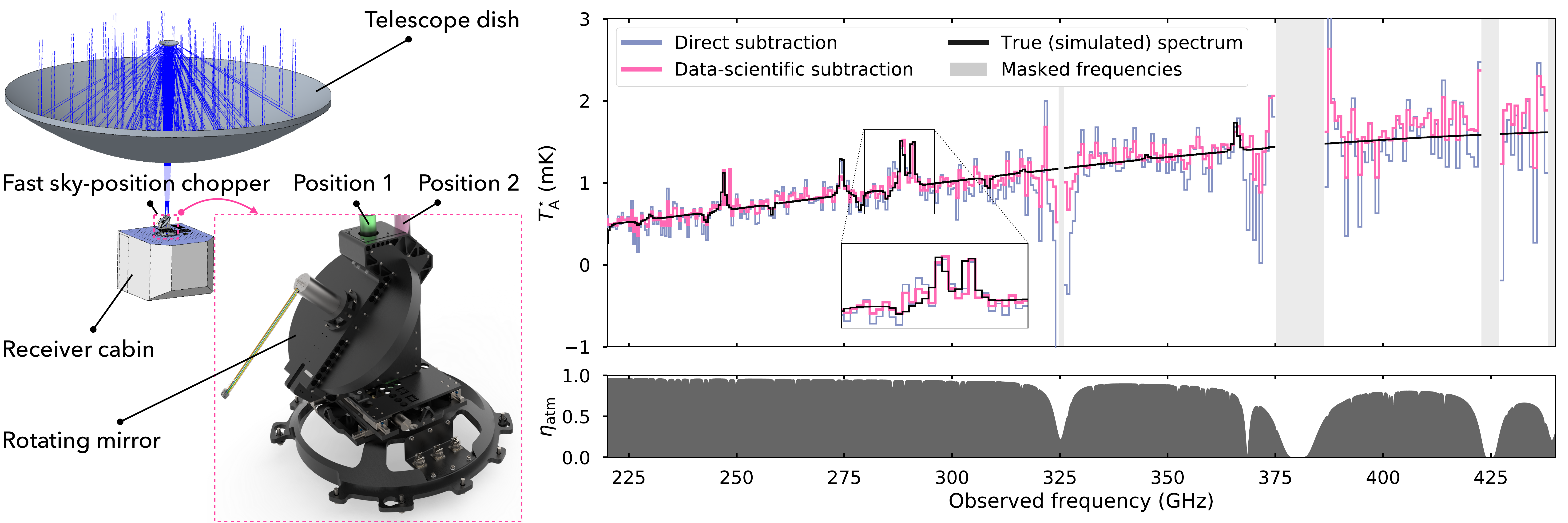}
    \caption{
        Illustration of the fast sky-position chopper for ASTE (left panel) and a simulation of a DESHIMA 2.0 observation with the chopper toward a $z=3$ DSFG reduced by different noise-removal methods (right panel).
        The simulated observation data including signals and noises are made by open-source codes \texttt{TiEMPO} \cite{Huijten_2022, tiempo} and \texttt{GalSpec} \cite{galspec}.
        The same conditions as Fig.~\ref{fig:detectable-line-flux} are assumed during a 30-min observation.
        In the direct subtraction method (a widely-used method), on- and off-source spectra of each switching cycle are subtracted in raw data form.
        In the data-scientific method (\texttt{SPLITTER} \cite{Brackenhoff_2021}), off-source spectra are modeled before subtraction, which avoids ``addition'' of stochastic photon noise or two-level noise of the instrument.
    }
    \label{fig:observation-strategy}
\end{figure*}

To make the observations more efficient, we adopt a data-scientific noise-removal approach \cite{Taniguchi_2021} for DESHIMA 2.0.
The stationary spectrum plus low-rank iterative transmittance estimator (\texttt{SPLITTER} \cite{Brackenhoff_2021}) decomposes observed time-series data (expressed as a matrix) into sky spectra (low-rank), astronomical signals (stationary), and stochastic noises without direct on-off subtraction.
Fig.~\ref{fig:observation-strategy} shows a simulation-based observation with DESHIMA 2.0 reduced by \texttt{SPLITTER}, where the line and continuum signals of a DSFG are successfully ``detected''.
The achieved noise level is improved by a factor of 1.7 compared to the direct on-off subtraction, which is close to the expected improvement factor of $\sqrt{2}$.
Testing this method with actual measured data is one of the objectives of the commissioning at ASTE.

%% file: science.tex
% !TEX root = ../manuscript.tex
\section{Science cases}
\label{s:science-cases}

We offer the single-pointing and on-the-fly mapping modes.
With the former mode enabled by the fast sky-position chopper, we plan a multi-line spectroscopic survey toward high-redshift DSFGs to identify their redshifts and investigate their physical properties.
The ultra-wideband capability of DESHIMA 2.0 will enable rapid measurements of, for example, [C~II] and highly-excited CO as probes of star-formation and interstellar heating sources such as AGNs, respectively.
The feasibility of the survey is studied in a separate paper \cite{Rybak_2022}.

With the latter mode, we plan a multi-frequency continuum mapping observation toward a galaxy cluster RX J1347.5-1145 to measure the spectral shape of the Sunyaev-Zel'dovich effect (SZE) signal and constrain the kinetic SZE component.
By combining several spectral channels, we estimate that a DESHIMA 2.0 observation with on-source time of eight hours will detect the signal in the 270~GHz band at S/N$\sim$5--10.
Note that, as ASTE/DESHIMA is a general-purpose spectrometer system, further science cases will also be possible.

%% file: conclusions.tex
% !TEX root = ../manuscript.tex
\section{Conclusions}
\label{s:conclusions}

We present the development of DESHIMA 2.0, an ultra-wideband (220--440~GHz) integrated superconducting spectrometer for submillimeter astronomy.
Various upgrades of the instrument design and the observation strategy studied in the laboratory fabrication, the measurement, and the simulation ensure that DESHIMA 2.0 will detect faint emission from a high-redshift galaxy overnight, enabling the coming scientific observation campaign with the ASTE 10-m telescope in 2023.

%% file: backmatter.tex
% !TEX root = ../manuscript.tex
\begin{acknowledgements}
    This research was supported by the Japan Society for the Promotion of Science JSPS (KAKENHI grant Nos. 17H06130, 18K03704) and the Joint Research Program of the Institute of Low Temperature Science, Hokkaido University (grant Nos. 21G024, 20G033).
    AE was supported by the Netherlands Organization for Scientific Research NWO (Vidi grant No. 639.042.423).
    JJAB was supported by the European Research Council ERC (ERC-CoG-2014 - Proposal n$^\circ$ 648135 MOSAIC).
    TT was supported by MEXT Leading Initiative for Excellent Young Researchers (grant No. JPMXS0320200188).
    YT and TJLCB are supported by NAOJ ALMA Scientific Research (grant No. 2018-09B).
    The ASTE telescope is operated by the National Astronomical Observatory of Japan (NAOJ).
\end{acknowledgements}

\bibliographystyle{spphys}
\bibliography{references}